\documentclass[final,1p,times]{elsarticle}

\usepackage{amsmath}
\usepackage{amssymb}
\usepackage{epsfig}

\newcommand* {\ket}[1]{\ensuremath{| {#1} \rangle}}
\newcommand* {\ee}{\ensuremath{\mathrm{e}}}

\journal{Physica B}

\begin{document}

\begin{frontmatter}

\title{Optical conductivity of single-layer graphene induced by temporal mass-gap fluctuations}

\author{J\'ozsef Zsolt Bern\'ad}

\ead{Zsolt.Bernad@physik.tu-darmstadt.de}

\address{Institut f\"{u}r Angewandte Physik, Technische Universit\"{a}t Darmstadt,
D-64289, Germany}
\address{Institute of Fundamental Sciences and MacDiarmid Institute for Advanced
Materials and Nanotechnology, Massey University, Private Bag 11~222, Palmerston
North 4442, New Zealand}

\begin{abstract}
We consider the dynamics of charge carriers in single-layer graphene that are subject
to random temporal fluctuations of their mass gap. The optical conductivity is calculated
by incorporating the quantum-stochastic time evolution into the standard linear-response
(Kubo) theory. We find that, for an intermediate range of frequencies below the average
gap size, electron transport is enhanced by fluctuations. At the same time, in the limit of
high as well as low frequencies, the conductivity is suppressed as the variance of gap
fluctuations increases. In particular, the dc conductivity is always suppressed by a
random temporal mass with nonvanishing mean value and vanishes in the zero-temperature
limit. Our results are complementary to those obtained recently for static random-gap disorder
in finite-size systems.
\end{abstract}


\end{frontmatter}

\section{Introduction}

Graphene is a newly accessible nanomaterial~\cite{Nov1,Nov2,Geim1,Geim2}, consisting
of a single sheet of carbon atoms forming a two-dimensional honeycomb lattice. The valence
(highest occupied) and conduction (lowest unoccupied) bands of graphene are touching
at the $\mathbf K$, $\mathbf K^\prime = -\mathbf K$ (and equivalent) points in the Brillouin
zone and exhibit a conical shape of their dispersion in the vicinity of these special (Dirac)
points~\cite{Wallace,Slonczewski,Semenoff,castroneto}. Graphene is a promising candidate
for applications in future micro- and nanoelectronics due to its excellent mechanical 
characteristics, scalability to nanometer sizes, and the ability to sustain huge electric
currents~\cite{Morozov,Chen}. Recent experiments have confirmed that the charge carriers in
graphene indeed behave like massless Dirac fermions~\cite{Nov1, YZhang}.

Although graphene exhibits excellent conducting properties, the absence of an energy gap
poses a challenge for realizing conventional semiconductor device operations in this material.
A possible way to alleviate this situation is to induce a gap in the electronic spectrum at the
Fermi energy by breaking the discrete sublattice symmetry of the honeycomb structure.
In recent experiments~\cite{elias, bostwick} was reported that hydrogenation of the graphene sheet breaks the sublattice degeneracy.   
Reversibility of this process
enables switching between conducting and insulating regimes of a graphene sample, and
spatial addressability should make the fabrication of hybrid conducting-insulating graphene
samples possible.

The opening of a uniform gap destroys graphene's metallic state, as the ac (optical)
conductivity vanishes for photon energies smaller than the band gap. In real samples,
the gap may be random, e.g., because of structural variability in the physical system. 
For example, the bonds with hydrogen atoms
could undergo temporal and spatial variations. To understand this practical issue, and
also for more fundamental reasons, studies of {\em static\/} random-gap disorder in
Dirac-fermion systems have been performed~\cite{Ludwig, Ziegler, Badarson, sinner}, where a staggered potential
was considered as a possible model for breaking the sublattice symmetry of the honeycomb lattice.
Our work presented here augments these previous investigations by considering the
effect of a spatially uniform mass gap that fluctuates {\em in time\/}. Besides clarifying the
respective nature of temporal and spatial randomness in the mass gap, which is an issue
of basic interest, the scenario considered by us would be realized in graphene samples
where sublattice-potential-inducing agents are undergoing large-scale random variations.
Below we argue that a more general spatio-temporal random variation of a staggered
sublattice potential in graphene can be represented, to leading order, by a Markovian
model where the low-energy Dirac-fermion theory has a random mass gap described by
white-noise fluctuations. We use this model to derive our stochastic equation of motion for
the disordered-graphene system and incorporate the nontrivial time evolution into the
framework of linear-response theory. Thus a generalized Kubo formula is obtained for
the optical conductivity, enabling us to unambiguously determine how electric transport
is affected by the mean value and the variance of the fluctuating potential. This is a
useful result because the optical conductivity is an experimentally accessible
quantity~\cite{Kuzmenko,Mak,Wang,Li} that depends strongly on the electronic properties
of the material. In particular, it can be used to determine the size of a band gap~\cite{Gusynin}
since transport is suppressed for photon energies smaller than the gap.
We find that the dc conductivity at finite temperature is monotonously decreasing as a
function of the variance of mass-gap fluctuations. However, for an intermediate range of
frequency below the cut-off equivalent to the average gap, fluctuations {\em enhance\/} the
conductivity. This observation should be contrasted with the related increase of the conductivity
with variance of {\em static\/} mass-gap fluctuations in graphene samples of finite
size~\cite{Ziegler, sinner}.

The remainder of this paper is organized as follows. Our theoretical method is outlined in the
following Sec.~\ref{sec:theory}, where a generalized linear-response formula for the conductivity
of a system subject to random temporal fluctuations is derived. Results obtained from application
of this formalism to randomly gapped graphene are presented in Sec.~\ref{sec:results}. 
We explore dependencies on various experimentally controllable parameters and discuss
the relation of our results to previous work. Our conclusions are given in Sec.~\ref{sec:concl}. 

\section{Theoretical method}
\label{sec:theory}

\subsection{Fluctuating gap model}
Quantum (and classical) systems
experience dissipation and fluctuations through interaction with a reservoir or environment~\cite{motiv}. 
Here, the hydrogenization of the graphene sample induces states, which we consider as a reservoir. 
The interaction with the states of the reservoir generates fluctuation, and a short explanation of the technical details 
is given in \ref{app}. 
In the context of the open quantum system (subsystem and reservoir), we consider the tight-binding Hamiltonian for electrons in 
graphene subject to a sublattice-staggered potential,
\begin{eqnarray}
 \hat{\mathcal{H}}=-t \sum_{\langle i,j \rangle, s} \left ( \hat{a}^\dagger_{i,s}\hat{b}_{j,s} 
+H.C. \right)+\sum_{i,s}V_{i,a}\, \hat{a}^\dagger_{i,s} \hat{a}_{i,s} 
+\sum_{i,s}V_{i,b}\, \hat{b}^\dagger_{i,s} \hat{b}_{i,s},
\end{eqnarray}
where $\hat{a}^\dagger_{i,s}$, $\hat{a}_{i,s}$ annihilates (creates) an electron with
spin $s$ ($s=\uparrow,\downarrow$) on site $i$ on sublattice $A$ (͑an equivalent 
definition is used for $\hat{b}^\dagger_{i,s}$, $\hat{b}_{i,s}$ on sublattice $B$), $t͑ \approx 2.8$eV is the
nearest-neighbor hopping energy between different sublattices. $V_{i,a}$ and $V_{i,b}$ are spin-independent potentials
with $V_{i,a}=m_i$ on sublattice $A$ and $V_{i,b}=-m_i$ on sublattice $B$. These potentials break the sublattice symmetry
of the single-layer graphene.

As a consequence of the interaction with the reservoir, $m_i$ is a time-dependent random variable at lattice site $i$, 
having a mean value $\langle
\langle m_i(t) \rangle \rangle=\bar{m}$ and a variance $\langle \langle (m_i(t)-\bar{m})(m_j(t')
-\bar{m}) \rangle \rangle=g^2 \delta_{i,j}\delta(t-t')$. 
The double expectation value is just the average over the local probability distribution
function related to the lattice site $i$, followed by averaging over the whole lattice probability
distribution function. In a more general case~\cite{example}, values for $m_i(t)$
at different lattice sites could be correlated. We make the assumption that the random variables 
are uncorrelated for different lattice sites and the value of the average gap (averaging locally) is
the same for each lattice site.
This random-mass variable can be modeled as $m_i(t)=\bar{m}+g \xi_i(t)$, where $\xi_i(t)$ represents temporal
white-noise fluctuation at lattice site $i$, and $\langle \langle \xi_i(t) \xi_j(t') \rangle \rangle=
\delta_{i,j}\delta(t-t')$.

We remind the reader that the static disorder models discuss only spatial correlations between different lattice sites.
The aforementioned description given by us could deal with local (temporal fluctuations) and spatial correlations, too. 
However, in the present paper we consider a complementary case to the static disorder models. Our model for randomness is given 
by the relation $m_i(t)=\bar{m}+g \xi_i(t)$, and it is clear that an average over the local fluctuations 
leads to a spatially uncorrelated variable $\bar{m}$.

The usual derivation \cite{Semenoff} of the low-energy continuum model for quasiparticles in
graphene yields the Hamiltonian
\begin{equation}
\label{model}
\hat{H}(\mathbf{k})=\hbar v (\sigma_x \hat{\mathbf{k}}_x+\sigma_y \hat{\mathbf{k}}_y)+m(t) \sigma_z \hat{\mathbf{1}}_{\mathbf{k}},
\end{equation}  
with multiplication operators $\hat{\mathbf{k}}_x, \hat{\mathbf{k}}_y, \hat{\mathbf{1}}_{\mathbf{k}}$
(that become numbers $k_x, k_y, 1$ in the plane-wave representation), and $\sigma_i$
denoting the Pauli matrices for the pseudo-spin degree of freedom. $v$ is the Fermi velocity,
which has a value $\simeq 10^6$~m/s. The $m(t)$ represents the randomness introduced
in the tight-binding model; it is equal to $\bar{m}+g \xi(t)$, where $\xi(t)$
is the usual white noise. 
The white noise, $\xi(t)$, dependence on the $\xi_j(t)$ is defined by $\sum_j \xi_j(t) \ee^{i({\mathbf k}'-{\mathbf k})\cdot
{\mathbf R}_j}=\xi(t)\delta({\mathbf k}'-{\mathbf k})$, where ${\mathbf k}$, ${\mathbf k}'$ are the
variables of the Fourier transformation, and ${\mathbf R}_j$ is the position vector of lattice site
$j$. \\
We assume disorder potentials to be weak (inducing small momentum scattering) such that the
inequivalent Dirac points associated with the ${\mathbf K}$ and ${\mathbf K}^\prime$
valleys in graphene's band structure are uncoupled~\cite{mccann} and, thus, contribute
additively to electronic transport. This allows us to consider the presence of randomness for 
each valley separately. The fact that electrons in graphene also carry a real spin introduces
an additional double degeneracy of all eigenvalues for the Hamiltonian (\ref{model}). In the
following, we study the dynamics due to the single-valley, spin-less model defined by
Eq.~(\ref{model}) and absorb degeneracy factors into our basic unit of conductivity.

\subsection{Quantum-stochastic time evolution for the density matrix}

As the Hamiltonian (\ref{model}) contains a random variable, the time evolution induced by it
cannot be expressed by the usual equations. Instead, a quantum-stochastic calculus has to
applied. A brief introduction to that is presented in the following.~\cite{Parthasarathy,Petruccione}
An isolated quantum system evolves in a unitary fashion. A physical quantity that is given at
time $t=0$ by an observable $\hat{A}$, will be described at time $t>0$ by $\hat{A}(t)=
\hat{U}^\dagger(t) \hat{A}\hat{U}(t)$, where $\hat{U}(t)$ is a unitary operator for 
each point in time $t$. The unitary operator is generated by the Schr\"odinger equation:
\begin{equation}
\frac{d \hat{U}(t)}{dt}=-\frac{i}{\hbar}\hat{H}_0(t)\hat{U}(t),
\end{equation}
where the (time dependent) Hamiltonian $\hat{H}_0(t)$ is a self-adjoint operator for each $t$. 
The Schr\"odinger equation tells us that in a short time interval $dt$, the unitary operator
changes at time $t$ like
\begin{equation}
d\hat{U}(t)=-\frac{i}{\hbar}\hat{H}_0(t)\hat{U}(t) dt.
\end{equation}
The Hamiltonian in Eq.~\eqref{model} can be written in the form of $\hat{H}(\mathbf{k})=
\hat{H}_0(\mathbf{k})+\hat{H}_1 \xi(t)$, where
\begin{subequations}
\begin{eqnarray}
\hat{H}_0(\mathbf{k})&=&\hbar v (\sigma_x \hat{\mathbf{k}}_x+\sigma_y \hat{\mathbf{k}}_y)
+\bar{m} \sigma_z \hat{\mathbf{1}}_{\mathbf{k}},\\
\hat{H}_1&=&g\sigma_z \hat{\mathbf{1}}_{\mathbf{k}} .
\end{eqnarray}
\end{subequations}
The white noise $\xi(t)$ is the formal derivative of a Wiener process $W_t$ (continuous everywhere but differentiable nowhere), 
a Gaussian random variable with zero mean value
\begin{equation}
 \mathbf{M}(W_t)=0,
\end{equation}
and variance $t$
\begin{equation}
 \mathbf{M}(W^2_t)-\mathbf{M}(W_t)^2=t.
\end{equation}
The stochastic calculus applied here will be based on the Wiener process $W_t$.
The above properties are represented in the differential equations as
\begin{equation}
\mathbf{M}(dW_t)=0,\,\,d^2W_t=dt,\,\,d^nW_t=0, \,n>2, 
\end{equation}
all of which will be applied in further calculations. Consider a stochastic processes $X_t$ governed by a stochastic differential
equations
\begin{equation}
dX_t=V_x dt + D_x dW_t, \nonumber
\end{equation}
where physical meaning of $V_x$ is the drift and of $D_x$ is the diffusion.
The Ito rule for a smooth function $f(x)$ states:
\begin{eqnarray}
df(X_t)&=&f'(X_t) dX_t + \frac{1}{2} f''(X_t) (dX_t)^2, \nonumber \\
f'(x)&=&\frac{df(x)}{dx},\,\, f''(x)=\frac{d^2f(x)}{dx^2},\,\, x \in \mathbb{R}, \nonumber \\
(dX_t)^2&=&D^2_x dt. \nonumber 
\end{eqnarray}
The quantum Ito rule~\cite{HudsonP} is the same as the Ito rule, only the noncommutativity of the operators is considered in addition.

Now, we apply the properties of the stochastic calculus, and we see that the unitary operator changes within an 
infinitesimal short time interval $dt$ at time $t$ as follows:
\begin{equation}
d\hat{U}(t)=\left(-\frac{i}{\hbar}\hat{H}_0(\mathbf{k})dt-\frac{1}{2 \hbar^2}\hat{H}^2_1 dt -
\frac{i}{\hbar}\hat{H}_1 dW_t  \right) \hat{U}(t).
\end{equation}

The equation of motion of the density matrix is given by $\hat{\rho}(t)=\hat{U}(t)\hat{\rho}
\hat{U}^\dagger(t)$, and a straightforward calculation yields
\begin{eqnarray}
\label{density}
d\hat{\rho}(t)=-\frac{i}{\hbar}[\hat{H}_0(\mathbf{k}),\hat{\rho}(t)] dt-\frac{1}{2\hbar^2}[\hat{H}_1,[\hat{H}_1,
\hat{\rho}(t)]]dt
-\frac{i}{\hbar}[\hat{H}_1,\hat{\rho}(t)] dW_t.
\end{eqnarray}
The current operator in our system is defined by $\hat{j}_{\mu}=e \frac{d\hat{r}_{\mu}}{dt}$,
where the time evolution of the coordinate operator satisfies
\begin{eqnarray}
\label{coordinate}
d\hat{r}_{\mu}(t)=\frac{i}{\hbar}[\hat{H}_0(\mathbf{k}),\hat{r}_{\mu}(t)] dt-\frac{1}{2\hbar^2}[\hat{H}_1,[\hat{H}_1,
\hat{r}_{\mu}(t)]]dt 
+\frac{i}{\hbar}[\hat{H}_1,\hat{r}_{\mu}(t)] dW_t.
\end{eqnarray}
$\hat{r}_{\mu}$ is a derivative operator in the $\mathbf{k}$-space and, in the sublattice representation, is given by $\hat{\mathbf{1}}_2\otimes i \frac{\partial}{\partial k_{\mu}}$.
This operator commutes with $\hat{H}_1$, which has a structure of $\sigma_z \otimes \hat
{\mathbf{1}}_{\mathbf{k}}$. These conditions and Eq.~\eqref{coordinate} define the current
operator as $\hat{j}_{\mu}=\frac{ie}{\hbar}[\hat{H}_0(\mathbf{k}),\hat{r}_{\mu}]=\frac{e}{\hbar}
\frac{\partial\hat{H}_0( \mathbf{k})}{\partial k_{\mu}}$. Due to the operator structure in
Eq.~\eqref{model}, the single-particle eigenstates $\ket{n}$ of this model can be written as a
direct product of a plane wave in configuration space with a spinor: $\ket{n}=\ket{\mathbf{k}} 
\otimes\ket{\pm}_{\mathbf{k}}$ \cite{castroneto,Bernad}. Here $\pm$ labels the electron
and hole bands, respectively, and the spinor wave function depends on wave vector
$\mathbf{k}$. The Cartesian current components are $\hat{j}_x= e v \sigma_x$ and
$\hat{j}_y= e v \sigma_y$ \cite{superselection}. The definition of the equilibrium density matrix
in the spinor space is $\hat{\rho}_0 \ket{\pm}_{\mathbf{k}}=f( \epsilon_{\mathbf{k},\pm})
\ket{\sigma}_{\mathbf{k}}$, and $\hat{H}_0 \ket{\pm}_{\mathbf{k}}= \epsilon_{\mathbf{k},
\pm}\ket{\pm}_{\mathbf{k}}$, where $f$ is the Fermi-Dirac distribution function and
$\epsilon_{\mathbf{k},\pm}=\pm \sqrt{(\hbar v)^2|\mathbf{k}|^2+\bar{m}^2}$. This equilibrium
state is the average of all equilibrium realizations defined by $\hat{\rho}_0=\ee^{-\beta \hat{H}
(\mathbf{k})}/{\mathrm{Tr}}(\ee^{-\beta \hat{H}(\mathbf{k})})$.

\subsection{Generalization of linear-response formalism}

We employ the linear-response (Kubo) formalism~\cite{Madelung} and
divide the system's Hamiltonian into the part $\hat{H}(\mathbf{k})$, which governs the 
evolution in Eq. \eqref{density}, and $\delta\hat{H}$, the perturbation associated with an external electric
field $\mathbf{E}$. For simplicity, we take the latter to be constant in space and
assume the field to be applied between $t=-\infty$ and $t=0$. The perturbation
Hamiltonian is $\delta \hat{H}=-e \mathbf{E}\cdot \hat{\mathbf{r}}\,\ee^{i \omega t}$. The equation of motion of the 
system with the added external field is:
\begin{eqnarray}
\label{linearmaster}
d\hat{\rho}(t)=&&-\frac{i}{\hbar}[\hat{H}_0(\mathbf{k}),\hat{\rho}(t)] dt-\frac{i}{\hbar}[\delta \hat{H},\hat{\rho}(t)]dt  
-\frac{1}{2\hbar^2}[\hat{H}_1,[\hat{H}_1,\hat{\rho}(t)]]dt
-\frac{i}{\hbar}[\hat{H}_1,\hat{\rho}(t)] dW_t \nonumber \\
=&& \mathcal{L} \hat{\rho}(t) dt-\frac{i}{\hbar}[\delta \hat{H},\hat{\rho}(t)]dt-\frac{i}{\hbar}[\hat{H}_1,\hat{\rho}(t)] dW_t.
 \nonumber
\end{eqnarray}
Within linear-response theory, we can linearize $\hat{\rho}=\hat{\rho}_0+
\delta{\hat{\rho}}$, where $\hat{\rho}_0$ is the system's equilibrium density matrix. Keeping
only linear terms in Eq.~\eqref{linearmaster}, we get
\begin{eqnarray}
d \delta{\hat{\rho}}=\mathcal{L}\delta{\hat{\rho}} dt -\frac{i}{\hbar}[\delta \hat{H},
\hat{\rho}_0] dt -\frac{1}{2\hbar^2}[\hat{H}_1,[\hat{H}_1,\hat{\rho}_0]] dt 
-\frac{i}{\hbar}[\hat{H}_1,\hat{\rho}_0+\delta{\hat{\rho}}] dW_t,
\end{eqnarray} 
with using
$[\hat{H}_0(\mathbf{k}),\hat{\rho}_0]=0$ as well as $[\delta \hat{H},\delta \hat{\rho}]\simeq 0$.
Introducing $\Delta \hat{\rho}=\ee^{-\mathcal{L}t}\delta \hat{\rho}$ yields
\begin{eqnarray}
 d \Delta \hat{\rho}=\ee^{-\mathcal{L}t}\left(-\frac{i}{\hbar}[\delta 
\hat{H},\hat{\rho}_0]-\frac{1}{2\hbar^2}[\hat{H}_1,[\hat{H}_1,\hat{\rho}_0]]\right)dt 
- \frac{i}{\hbar} \ee^{-\mathcal{L}t} \left([\hat{H}_1, \hat{\rho}_0 + \ee^{\mathcal{L}t}\Delta \hat{\rho}]\right)dW_t.
\end{eqnarray}
Note that $\Delta \hat{\rho}$ and $\delta \hat{\rho}$ have the same value at $t=0$,
and both vanish at $t=-\infty$. Integration yields
\begin{equation}\label{innerres}
\delta \hat{\rho}(t=0)=\int^{0}_{-\infty}dt \,\,\ee^{-\mathcal{L}t}\left(-\frac{i}{\hbar}
[\delta \hat{H},\hat{\rho}_0]-\frac{1}{2\hbar}[\hat{H}_1,[\hat{H}_1,\hat{\rho}_0]]\right) +
\lim_{t \to \infty}\int^{0}_{-t} dW_t \left(\ee^{-\mathcal{L}t}\left(\frac{i}{\hbar}[\hat{H}_1,
\hat{\rho}_0] \right)-\frac{i}{\hbar}[\hat{H}_1, \Delta \hat{\rho}] \right).
\end{equation}
The exponential factor in Eq.~(\ref{innerres}) ensures convergence of the time 
integral, making it unnecessary to introduce the phenomenological adiabatic
damping parameter employed in conventional linear-response
theory~\cite{Madelung}. 

\subsection{Generalized conductivity formula}

We use (\ref{innerres}) to calculate the double expectation value (over the basis of the
one-particle Hilbert space and over the ensemble of realizations for the stochastic
process)
for the current operator $\hat{j}$ and use the property $\mathbf{M}(dW_t)=0$. 
We remind the reader that our stochastic variable models the local gap fluctuations. An average 
over this randomness yields the mean conductivity~\cite{extension} that is typically calculated and is our quantity
of interest here.
Dividing the
current expectation value by $|\mathbf{E}|$ yields the optical conductivity 
\begin{equation} \label{linearmaster2}
 \sigma_{\mu \nu}(\omega)=\int_{-\infty}^{0} \left[ K_{\mu\nu}(t) e^{i \omega t} +
L_{\mu\nu}(t) \right]\, dt \, , 
\end{equation}
with the kernels
\begin{equation}\label{KKp}
K_{\mu\nu}(t)=-\frac{1}{i\hbar} {\mathrm{Tr}}\left\{ \hat{j}_{\mu} \ee^{-\mathcal{L}t}\left( [e\hat{r}_{\nu},\hat{\rho}_{0}]
\right)\right\} \, ,
\end{equation}
\begin{equation}\label{KKp2}
L_{\mu\nu}(t)= {\mathrm{Tr}} \left \{\frac{\hat{j}_{\mu}}{E_{\nu}} \ee^{-\mathcal{L}t}\left(-\frac{1}{2\hbar^2}[\hat{H}_1,[
\hat{H}_1,\hat{\rho}_0]]\right)
\right\}\, .
\end{equation}
The ${\mathrm{Tr}}$ symbol stands for taking the trace over $2\times2$ matrices and
performing the integration $\int d^2 \mathbf{k}$. As the integrand in Eq.~\eqref{KKp2} is an odd
function of both $k_x$ and $k_y$, the integration over the polar angle yields zero and, hence,
the quantity $L_{\mu\nu}(t)$ vanishes. 

We remind the reader that the current flowing through the system was not affected by the fluctuations, see Eq. \eqref{coordinate} and 
the discussion afterwards. Otherwise, the variance of the fluctuations must be added to the conductivity formula. This constant would shift
the universal conductance of the graphene sheet, and would refute the experimental results. This is not the situation here, which shows the 
correctness of our model.

The calculation of $K_{\mu\nu}$ is straightforward, and
using the Laplace transform to solve for the dynamics, we find
\begin{equation}
K_{\mu\nu}(t)=\frac{e^2}{\hbar}\int \frac{d^2 \mathbf{k}}{(2 \pi)^2} \mathbf{Res}\left\{
\frac{N(\mathbf{k},\bar{m},g,z)}{D(\mathbf{k},\bar{m},g,z)}e^{zt}\right\},
\end{equation}
where $\mathbf{Res}$ stands for the sum of residues of the integrand. The conductivities related to the kernels $K_{xx}$ and 
$K_{yy}$ are identical, and we present the effect of the fluctuating gap for these particular conductivities. We have to mention
that the Hall conductivities $\sigma_{xy}$ and $\sigma_{yx}$ are equal to zero, even in the presence of the fluctuations. 
Without an external magnetic field, the 
contribution of the two valleys of the honeycomb-lattice band-structure cancel each other. 

Introducing polar coordinates for $\mathbf{k}$ and performing the angular integration yields 
\begin{equation}\label{K}
K_{\mu\nu}(t)=\frac{e^2}{\hbar}\int_0^{\infty} \frac{|\mathbf{k}| \,\,d|\mathbf{k}|}{4 \pi} \mathbf{Res}
\left\{\frac{N(|\mathbf{k}|,\bar{m},g,z)}{D(|\mathbf{k}|,\bar{m},g,z)}e^{zt}\right\} .
\end{equation}
For simplicity, the Fermi velocity $v$ has been absorbed into $|\mathbf{k}|$, and we find
\begin{eqnarray}
N(|\mathbf{k}|,\bar{m},g,z)&=&|\mathbf{k}|^2 \sqrt{m^2 + |\mathbf{k}|^2} \left(4 m^2+
4 |\mathbf{k}|^2 +z^2-4 \Gamma z \right) \times\sum \left(-
\frac{df( \epsilon)}{d\epsilon}\mid_{\epsilon=\epsilon_{\mathbf{k},\pm}} \right)\nonumber \\
&+&z (z-4 \Gamma) (2 m^2+|\mathbf{k}|^2) \left[ f(\epsilon_{\mathbf{k},-})-f(\epsilon_{\mathbf{k},+}) \right]\, , 
\end{eqnarray}
\begin{eqnarray}\label{denominator}
D(|\mathbf{k}|,\bar{m},g,z)= ( m^2 + |\mathbf{k}|^2 )^\frac{3}{2} \left[ 16 z \Gamma^2-8 (2|\mathbf{k}|^2+z^2) \Gamma 
+ z (4m^2+4 |\mathbf{k}|^2 + z^2) \right],
\end{eqnarray} 
where the parameters $\Gamma=g^2/2 \hbar^2$ and $m=\bar{m}/\hbar$ have been introduced.
The cubic factor in the denominator (\ref{denominator}) has three roots $z_i$, which give the
poles in eq.~(\ref{K}).

\section{Optical conductivity of gapped graphene}
\label{sec:results}

For small $\Gamma$, the roots are to lowest order $z_1=0$ and $z_{2,3}= \pm 2 i
\sqrt{|\mathbf{k}|^2+m^2}$. Using this and performing the time-integration in the limit
$\Gamma\to 0$, the $intra$-band contribution
\begin{equation}\label{gap2}
\frac{\sigma}{\sigma_0}=\frac{\pi}{2}\delta(\omega)
\int_0^{\infty}\frac{|\mathbf{k}|^3}{ m^2 + |\mathbf{k}|^2 }\sum \left(-
\frac{df( \epsilon)}{d\epsilon}\mid_{\epsilon=\epsilon_{\mathbf{k},\pm}} \right)d|\mathbf{k}| ,
\end{equation}
and the $inter$-band contribution
\begin{equation}\label{gap}
\frac{\sigma(\omega)}{\sigma_0}=\frac{\pi}{8} \frac{\omega^2 +4m^2}{\omega^2}\frac{\sinh(\frac{\hbar \omega}
{2k_BT})}{\cosh(\frac{\mu}{k_BT})+\cosh(\frac{\hbar \omega}{2k_BT})} \Theta(\omega-2m),
\end{equation}
to the conductivity of the uniformly gapped single-layer graphene are found~\cite{Gusynin}.
The Dirac-delta peak in the $intra$-band conductivity is due to elastic transitions, which are
only possible at finite temperature and/or when the chemical potential is bigger than the
average gap. The scale factor $\sigma_0=4e^2/h$ accounts for spin and valley degeneracy.
%%%%%%%%%%%%%%%%%%%%%%%%%%%%%%%%%%%%%%%%%%%%%%%%%%%%%%%%%%%%%%%%%%%%%%%%%%%%%%%
\begin{figure}[t]
\begin{center}
\includegraphics[width=3.1in]{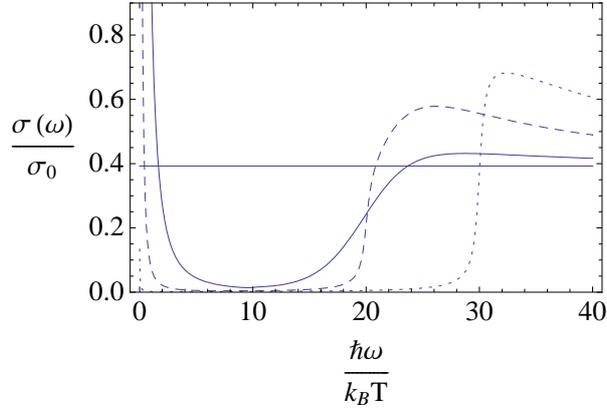}
\caption{\label{omega}
Optical conductivity of single-layer graphene with a weakly fluctuating mass gap. Results
shown are obtained for fixed chemical potential $\mu =10\,k_B T$ and gap-fluctuation variance
$\hbar \Gamma = 0.1\, k_B T$. The different curves correspond to $\bar{m}/ k_B T =5$ (solid),
$\bar{m}/ k_B T =10$ (dashed), and $\bar{m}/ k_B T =15$ (dotted). The value of  $\pi/8$ is
marked by a horizontal line. As can be seen, the ac conductivity exhibits a jump at twice the
average band-gap value $\bar{m}$  when the latter is greater than or equal to the chemical
potential.}
\end{center}
\end{figure}
%%%%%%%%%%%%%%%%%%%%%%%%%%%%%%%%%%%%%%%%%%%%%%%%%%%%%%%%%%%%%%%%%%%%%%%%%%%%%%%%%%

\begin{figure}[t]
\begin{center}
\includegraphics[width=3.1in]{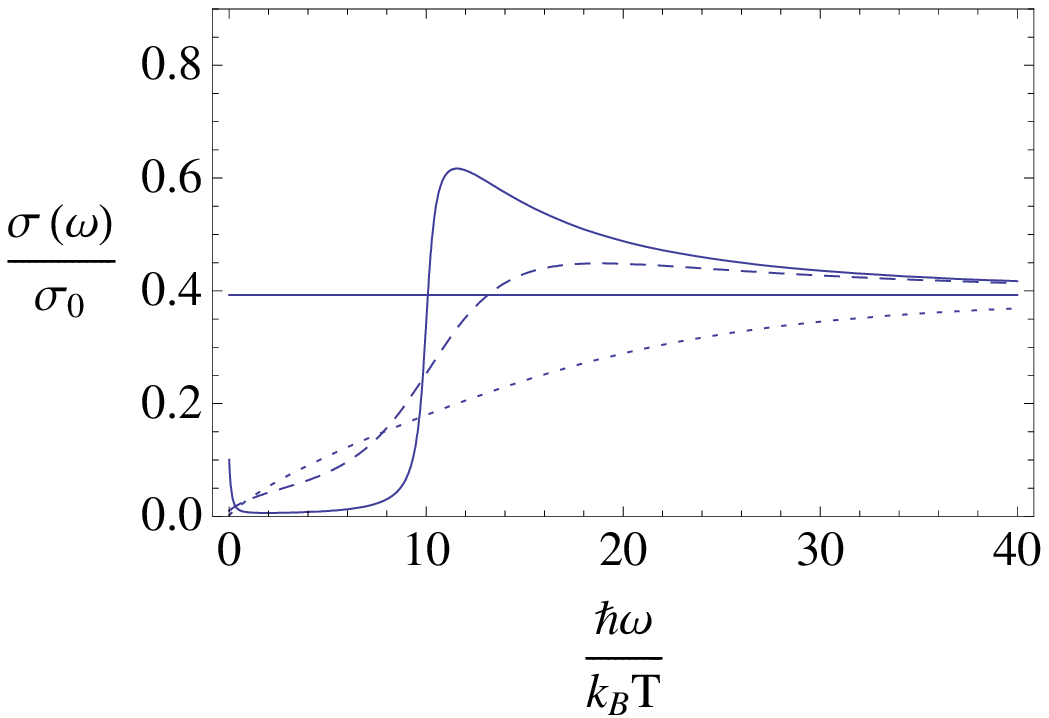}
\includegraphics[width=3.1in]{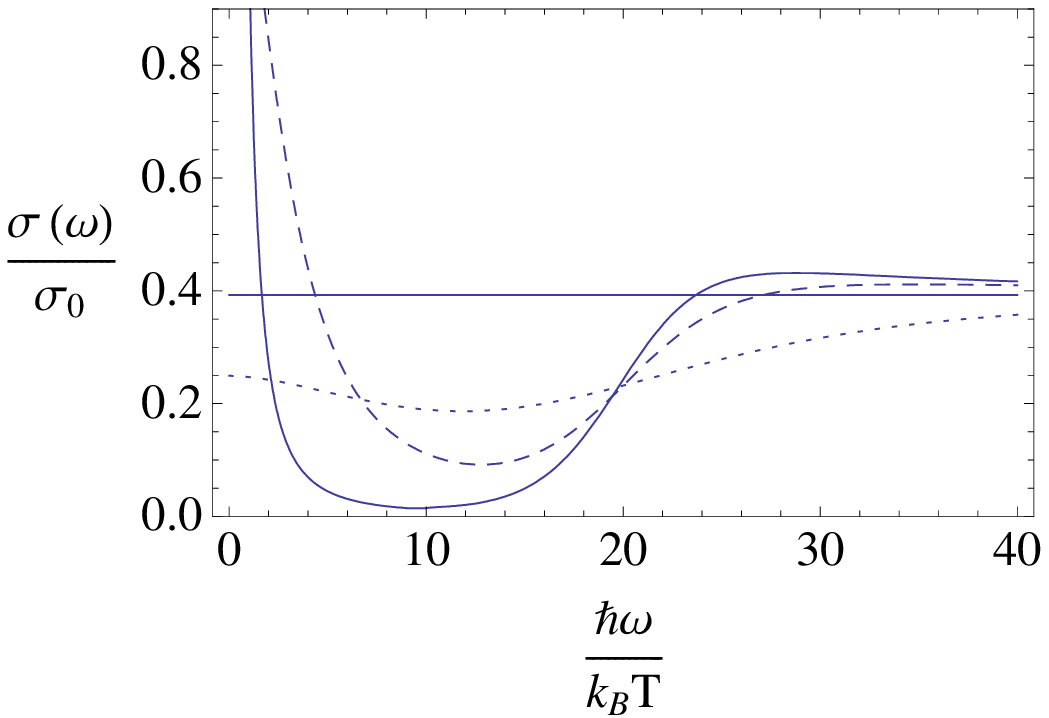}
\caption{\label{omega2}
Optical conductivity of single-layer graphene with a fluctuating gap. Top panel: the chemical
potential is fixed at the neutrality point ($\mu/ k_B T =0$), and the average gap is $\bar{m}/
k_B T=5$. Bottom panel: $\mu=10\, k_B T$ and $\bar{m} =5 k_B T$. The different curves
correspond to $\hbar \Gamma/ k_B T =0.1$ (solid), $\hbar \Gamma/k_B T =1$ (dashed),
and $\hbar \Gamma/ k_B T =5$ (dotted). The value of $\pi/8$ is marked by the horizontal
line. As the value of $\Gamma$ is decreased, the curves resemble more closely the result
found for uniformly gapped graphene. Note the intermediate range of low frequencies where
an increase in randomness (i.e., variance of fluctuations) results in an enhanced conductivity.}
\end{center}
\end{figure}

\begin{figure}[t]
\begin{center}
\includegraphics[width=3.1in]{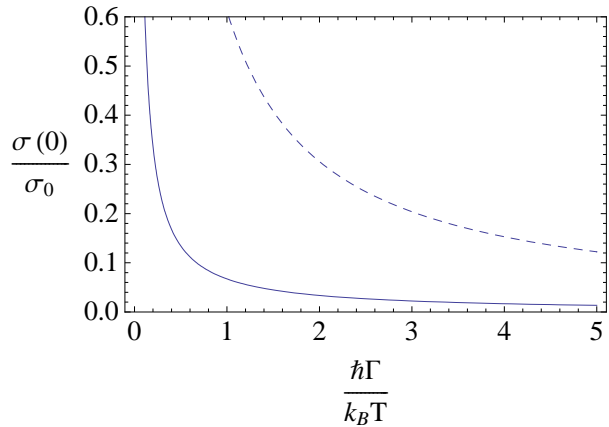}
\caption{\label{zero}The dc conductivity of randomly gapped graphene plotted as a function
of gap variance $\Gamma$. The average gap is fixed at $\bar{m}/ k_B T=2.5$. The 
different curves are for $\mu/ k_B T =0$ (solid) and $\mu/ k_B T =5$ (dashed). A decrease
in the dc conductivity is found for increasing $\Gamma$. This decrease is slower when the
chemical potential is bigger than the average gap. }
\end{center}
\end{figure}

\begin{figure}[t]
\begin{center}
\includegraphics[width=3.1in]{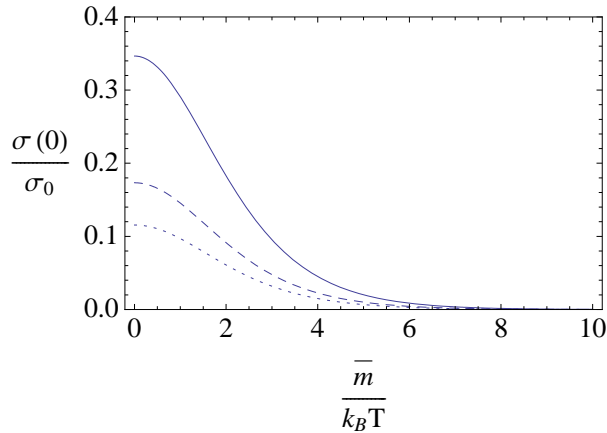}
\caption{\label{ma} The dc conductivity of randomly gapped graphene plotted as a function of
average gap for different values of gap variance $\Gamma$. The different curves correspond
to $\hbar \Gamma/ k_B T =0.5$ (solid), $\hbar \Gamma/ k_B T =1$ (dashed), and
 $\hbar \Gamma/ k_B T =1.5$ (dotted). The chemical potential is fixed at the neutrality point.
 As the average gap size is increased, the dc conductivity gets suppressed.}
\end{center}
\end{figure}

\begin{figure}[t]
\begin{center}
\includegraphics[width=3.1in]{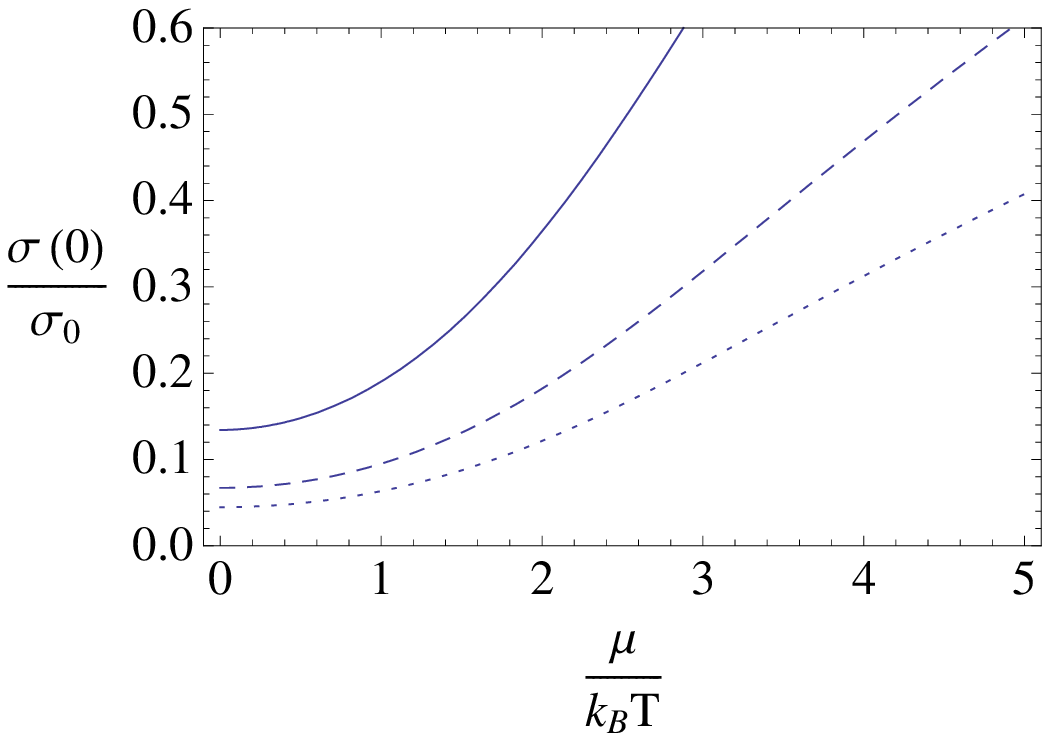}
\caption{\label{mu} The dc conductivity of randomly gapped graphene, plotted as a function
of chemical potential for different values of gap variance $\Gamma$. The different curves
were obtained for $\hbar \Gamma/ k_B T =0.5$ (solid), $\hbar \Gamma/ k_B T =1$ (dashed),
and $\hbar \Gamma/ k_B T =1.5$ (dotted). The average gap is fixed at $\bar{m}/ k_B T=2.5$.
The minimal value of the dc conductivity occurs at $\mu=0$ and decreases with increasing
$\Gamma$.}
\end{center}
\end{figure}

In the following, the conductivity is calculated numerically for finite values of
$\Gamma$ and $T$ from Eq.~(\ref{linearmaster2}) with Eq.~(\ref{K}). 
We use $k_B T$ as our unit of energy. Figs.~\ref{omega} and \ref{omega2} show
the ac (optical) conductivity, whereas Figs.~\ref{zero}, \ref{ma} and \ref{mu} show
results for the dc conductivity. 

%%%%%%%%%%%%%%%%%%%%%%%%%%%%%%%%%%%%%%%%%%%%%%%%%%%%%%%%%%%%%%%%%%%%%%%%%%%%%%%
%%%%%%%%%%%%%%%%%%%%%%%%%%%%%%%%%%%%%%%%%%%%%%%%%%%%%%%%%%%%%%%%%%%%%%%%%%%%%%%%%%
The interplay between average-gap size and chemical potential exhibits two regimes, as can
be seen by studying the uniform gapped model [result given by Eq.~\eqref{gap}] and also from
Fig.~\ref{omega}. The uniform gap model has a vanishing ac conductivity for photon energies
$\hbar \omega < 2 \bar{m}$, and a jump at $\hbar \omega = 2 \bar{m}$. The height of the 
jump depends on the chemical potential and the average gap. If the chemical potential is
much bigger than the average gap then we have a small jump height whereas, in the opposite
case, the jump is more noticeable. The ability to determine the average gap size by measuring
where the jump of the optical conductivity occurs depends also on the variance of the
fluctuations. In any case, setting the chemical potential as low as possible would be
favorable for the detection of the gap-induced jump in the ac conductivity.
 
Figure~\ref{omega2} shows the ac conductivity for a number of different variances of gap
fluctuations for two values of the chemical potential $\mu$: tuned to the neutrality point (top
panel) and for a large value of $\mu$ (bottom panel). While the situation with a small variance
of gap fluctuations is quite similar to the result found for a system with a uniform gap, the
increase of fluctuations generates sizable conduction in the frequency range $\omega <
2\bar m$. For both values of the chemical potential, a region emerges where an increase in
fluctuations results in an increased conductivity. Thus it appears that larger fluctuations will
facilitate electric transport in randomly gapped graphene sheets. A similar result has been
reported for static random-gap disorder in graphene~\cite{Ziegler, sinner}. However, in our
case, this behavior occurs only within a limited range of finite frequencies. The parametric
dependence on the fluctuation strength is reversed at low frequencies, in particular also for
the dc conductivity. For high frequencies, the conductivity approached the universal value
$\pi/8$. The detailed shape of the saturation depends on both $\Gamma$ and $\bar{m}$,
with higher values pushing convergence to higher frequencies.
%%%%%%%%%%%%%%%%%%%%%%%%%%%%%%%%%%%%%%%%%%%%%%%%%%%%%%%%%%%%%%%%%%%%%%%%%%%%%%%
%%%%%%%%%%%%%%%%%%%%%%%%%%%%%%%%%%%%%%%%%%%%%%%%%%%%%%%%%%%%%%%%%%%%%%%%%%%%%%%
%%%%%%%%%%%%%%%%%%%%%%%%%%%%%%%%%%%%%%%%%%%%%%%%%%%%%%%%%%%%%%%%%%%%%%%%%%%%%%%
%%%%%%%%%%%%%%%%%%%%%%%%%%%%%%%%%%%%%%%%%%%%%%%%%%%%%%%%%%%%%%%%%%%%%%%%%%%%%%%%%%
%%%%%%%%%%%%%%%%%%%%%%%%%%%%%%%%%%%%%%%%%%%%%%%%%%%%%%%%%%%%%%%%%%%%%%%%%%%%%%%
%%%%%%%%%%%%%%%%%%%%%%%%%%%%%%%%%%%%%%%%%%%%%%%%%%%%%%%%%%%%%%%%%%%%%%%%%%%%%%%%%%

The dc conductivity's dependence on the parameters $\Gamma$, $\bar{m}$, and the chemical 
potential is shown by Figs.~\ref{zero}, \ref{ma} and \ref{mu}. Increasing $\Gamma$ leads to a
decrease in the dc conductivity, which means that the $\Gamma$ simulate the same effect as
life-time broadening due to inelastic scattering. The rate of the decrease depends on the
relative magnitudes of $\mu$ and $\bar{m}$, as can be seen in Fig. \ref{zero}. If the chemical
potential is bigger than the average gap, the dc conductivity tends to zero slower with any
increase of $\Gamma$.  Increasing the average gap also leads to a decrease in the dc 
conductivity. The minimal value of the dc conductivity occurs at the charge neutrality point,
$\mu=0$, and strongly depends on the value of $\Gamma$ and $\bar{m}$.

Up until now, we considered electron transport at a finite temperature. To discuss the behavior
in the limit when $T \to 0$, we have to perform this limit already in Eq.~\eqref{K} and redo the
calculations following after that. In the case of $T \to 0$ and $\mu=0$ we are able to compare
our results to the works by Ziegler~\cite{Ziegler}. These works assume that the
static disorder average destroys the intra-band conductivity, which we also found for the above
assumptions. We are able to go further and in the case of $T \to 0$  and $\mu\leq  \bar{m}$ the
intra-band conductivity is still found to be zero. In the analytical works by Ziegler~\cite{Ziegler},
the inter-band conductivity contains a Heaviside function and has a contribution at $\omega=0$.
The role of the Heaviside function is to separate the insulating regime from the metallic one.
This is a point where we have a significant difference, because our Heaviside function is smeared by the $\Gamma$ parameter. In the limit $\omega=0$ with $\bar{m}\neq 0$ and
$\mu\leq  \bar{m}$, our inter-band conductivity is always zero.

The role played by the quantity $\Gamma$ in the present work is formally similar to a
decoherence parameter (also denoted by $\Gamma$ in a previous work~\cite{BJZ} by us)
measuring the effect of an ever-present environment. The parameter $\Gamma$ is the
coefficient of a double commutator that enters to the system's time evolution. In the present
case, the double commutator contains the $\hat{\sigma}_z$ operator, in contrast to our previous
work~\cite{BJZ} where it involved the $\hat{\sigma}_x$ and $\hat{\sigma}_y$ matrices. The
physical meaning of $\Gamma$ is also different: here it is related to the randomness of a
mass-gap-inducing degree of freedom such as hydrogen atoms, while it was used to describe
the properties of a current detector previously. We would also like to point out that, in general,
the gap fluctuations could be temperature-dependent, as most mechanisms for inducing a gap
(e.g., by structural modifications such as hydrogenation) will be affected by a variation of
temperature. As we have introduced the fluctuating gap in a phenomenological fashion, our
theory would apply to the ensemble of temporal fluctuations at a fixed temperature.

\section{Conclusions}
\label{sec:concl}

We have calculated the stochastic evolution of the density matrix for charge-carrier dynamics
in single-layer graphene subject to a fluctuating sublattice-staggered potential. We derived a
generalized Kubo formula to study the effect of temporal mass-gap fluctuations on the
conductivity. The variance of the fluctuations introduces a source of damping and thus makes
the converged adiabaticity parameter frequently used in Kubo formula calculations superfluous.
See also related work~\cite{BJZ}. Mixing of the $intra$-band and $inter$-band contributions to 
the ac and dc conductivities strongly affect its parametric dependence on the variance of the 
fluctuations $g^2=2\hbar^2 \Gamma$. A system with a uniform gap (i.e., vanishing variance
$\Gamma$) is insulating, as seen from Eq.~\eqref{gap}. For the more realistic case of a
fluctuating gap, a finite range of frequencies is found for which an increase in randomness
(i.e., variance of gap fluctuations) results in an increase of the ac conductivity. See
Fig.~\ref{omega2}. Such fluctuation-enhanced transport is not observed in the dc limit, where
increases in either the average size or the variance of the mass gap lead to a suppression of
conductivity at any finite temperature. In the zero-temperature limit, our model of randomly
gapped graphene exhibits insulating behavior when the average gap is
finite~\cite{Ludwig,Badarson} (but, as in the previous work~\cite{Ludwig}, a finite dc conductivity
is found when the average gap vanishes). 

The discussed differences between different random gap 
models can be found out in the properties of the conductivity, the only quantity retained after a measurement.\\  
 
The author has profited from helpful discussion with K.~Ziegler, A.~Sinner and U.~Z\"ulicke. This work
was supported by a postdoctoral fellowship grant from the Massey University Research Fund.
Additional funding through BMBF project QK{\_}QuOReP is gratefully acknowledged.

\appendix
 \section{Stochastic dynamics}
 \label{app}
A quantum master equation describes the dynamics of an open quantum system (subsystem+reservoir). The Markovian quantum master equation is 
given by a first-order differential equation for the density matrix of the subsystem,
\begin{equation}
\label{Leq}
 \frac{d}{dt}\hat{\rho}_S(t)=\mathcal{L} \hat{\rho}_S(t),
\end{equation}
where the most general form for the generator $\mathcal{L}$ is given by the Lindblad equation \cite{Lindblad,GKS},
\begin{eqnarray}
 \mathcal{L} \hat{\rho}_S=-\frac{i}{\hbar}[\hat{H}_S,\hat{\rho}_S]+\sum_k \gamma_k\left(\hat{A}_k \hat{\rho}_S \hat{A}^\dagger_k-
\frac{1}{2}\hat{A}^\dagger_k \hat{A}_k \hat{\rho}_S-\frac{1}{2} \hat{\rho}_S \hat{A}^\dagger_k \hat{A}_k \right).
\end{eqnarray}
The first term of the generator represents the unitary part of the dynamics generated by the Hamiltonian $\hat{H}_S$ of the subsystem. The 
operators $\hat{A}_k$ are the Lindblad operators and the quantities $\gamma_k$ have the dimension of an inverse time. The 
operators $\hat{A}_k$ and the quantities $\gamma_k$ are derived from the dynamics of the total system (subsystem and reservoir) in various
approximation schemes. The states of the reservoir are traced out during the procedure.
 
It is clear that the eq. \eqref{Leq} cannot be used for the derivation elaborated in the work of Semenoff \cite{Semenoff}. The master
equation for the subsystem can be reformulated in terms of a stochastic process for the subsystem's wave function. This idea is the 
so-called {\em unravelling\/} of the master equation \cite{Davies,BB}.

Now, we consider our  microscopic model of the total system,
\begin{eqnarray}
\label{Htot}
\hat{H}&=&\hat{H}^g+\hat{H}_I+\hat{H}^r, \nonumber \\
\hat{H}^g&=&-t \sum_{\langle i,j \rangle, s} \left ( \hat{a}^\dagger_{i,s}\hat{b}_{j,s} 
+H.C. \right)+\sum_{i,s}\bar{m}_i\, \hat{a}^\dagger_{i,s} \hat{a}_{i,s}
-\sum_{i,s}\bar{m}_i\, \hat{b}^\dagger_{i,s} \hat{b}_{i,s}, \nonumber \\
\hat{H}_I&=&\sum_{i,s}\lambda_{i}\, \hat{a}^\dagger_{i,s} \hat{a}_{i,s} \otimes \hat{H}^r_{I,i,s,a} 
-\sum_{i,s}\lambda_{i}\, \hat{b}^\dagger_{i,s} \hat{b}_{i,s} \otimes \hat{H}^r_{I,i,s,b}, 
\end{eqnarray}
where $\hat{a}^\dagger_{i,s}$, $\hat{a}_{i,s}$ annihilates (creates) an electron with
spin $s$ ($s=\uparrow,\downarrow$) on site $i$ on sublattice $A$ (͑an equivalent 
definition is used for $\hat{b}^\dagger_{i,s}$, $\hat{b}_{i,s}$ on sublattice $B$). $\hat{H}^r$ is the Hamiltonian of the states, considered
as reservoir. $\hat{H}_I$ is the interaction Hamiltonian and $\hat{H}^r_{I,i,s,a}$ and $\hat{H}^r_{I,i,s,b}$ are functions of creation 
and annihilation operators related to the states of the reservoir coupled to sublattices $A$ and $B$. The interaction Hamiltonian is chosen such that 
the sublattice symmetry is broken. $\lambda_{i}$ is a spin independent coupling
constant. $\bar{m}_i$ is the strength of the sublattice symmetry breaking on lattice site $i$.

The usual Born-Markov master equation \cite{Petruccione} can be derived by considering that the time scale of the hopping is 
much more faster than the relaxation of the reservoir states and the couplings are weak:
\begin{eqnarray}
\label{meq}
\frac{d\hat{\rho}_S}{dt}=\mathcal{L} \hat{\rho}(t)=-\frac{i}{\hbar}[\hat{H}^g,\hat{\rho}_S(t)] 
-\frac{1}{\hbar^2}\int_0^{t} {\mathrm{Tr}}_r\left[\hat{H}_{I}(t),\big[\hat{H}_{I}(s),\hat{\rho}_S(t)\otimes\hat{\rho}_r \big] \right] 
ds,  
\end{eqnarray}
where ${\mathrm{Tr}}_r$ stands for the partial trace over the reservoir states. $\hat{\rho}_r$ is the density matrix of the reservoir.
Inserting Eq. \eqref{Htot} into the master equation (Eq. \eqref{meq}) we obtain
\begin{eqnarray}
 \frac{d\hat{\rho}_S}{dt}&=&-\frac{i}{\hbar}[\hat{H}^g,\hat{\rho}_S(t)]-\frac{1}{\hbar^2} \sum_{i,s} \frac{g^2_i}{2} 
\left[\hat{H}^g_{I},\big[\hat{H}^g_{I},\hat{\rho}_S(t) \big] \right], \nonumber \\
\hat{H}^g_{I}&=& \hat{a}^\dagger_{i,s} \hat{a}_{i,s}- \hat{b}^\dagger_{i,s} \hat{b}_{i,s},
\end{eqnarray}
where $g_i$ is related to the reservoir correlation functions, and we considered that these correlations are the same for sublattice $A$
and $B$, too.

The {\em unravelling\/} of the above master equation gives the following stochastic Hamiltonian
\begin{eqnarray}
 \hat{H}=-t \sum_{\langle i,j \rangle, s} \left ( \hat{a}^\dagger_{i,s}\hat{b}_{j,s} 
+H.C. \right)+\sum_{i,s}\bar{m}_i(t)\, \hat{a}^\dagger_{i,s} \hat{a}_{i,s} 
-\sum_{i,s}\bar{m}_i(t)\, \hat{b}^\dagger_{i,s} \hat{b}_{i,s},
\end{eqnarray}
where
\begin{equation}
 \bar{m}_i(t)=\bar{m}_i+g_i \xi(t).
\end{equation}
$\xi(t)$ stands for the standard white noise.


\begin{thebibliography}{00}
\bibitem{example} 
As a simple example, but more general than the considered model in the main text, consider $m_i(t)$ given by $\bar{m}_i+g\xi_i(t)$:
averaging first locally, we get $\langle m_i(t) \rangle_i=\bar{m}_i$ and then for the whole lattice system
$\langle \bar{m}_i \rangle_N=\bar{m}$ is found. The variance $\langle (\bar{m}_i-\bar{m})(\bar{m}_j-\bar{m}) \rangle_N$ 
of the locally averaged variable $\bar{m}_i$ is nonzero.
\bibitem{superselection}
The Hilbert space is decomposed into irreducible components. Each pseudo-spin sector is a superselection sector, preserved by 
the current operator.
\bibitem{extension} A possible extension of this work is the full time dependence
involving the Wiener process, which gives a picture of the raw measured data obtained
for the conductivity.
\end{thebibliography}
\end{document}